\documentclass[12pt]{article}

\include{_preamble}
\AtEndDocument{\refstepcounter{theorem}\label{finalthm}}
\AtEndDocument{\refstepcounter{figure}\label{finalfig}}

\title{Non-parametric targeted Bayesian estimation of class
  proportions in unlabeled data}

\date{\today}

\begin{document}\maketitle

\begin{abstract}
  We introduce a novel Bayesian estimator for the class proportion in
  an unlabeled dataset, based on the targeted learning framework. Our
  procedure requires the specification of a prior (and outputs a
  posterior) only for the target of inference, instead of the prior
  (and posterior) on the full-data distribution employed by classical
  non-parametric Bayesian methods .When the scientific question can be
  characterized by a low-dimensional parameter functional, focus on
  such a prior and posterior distributions is more aligned with
  Bayesian subjectivism, compared to focus on entire data
  distributions. We prove a Bernstein-von Mises-type result for our
  proposed Bayesian procedure, which guarantees that the posterior
  distribution converges to the distribution of an efficient,
  asymptotically linear estimator. In particular, the posterior is
  Gaussian, doubly robust, and efficient in the limit, under the only
  assumption that certain nuisance parameters are estimated at slow
  rates. We perform numerical studies illustrating the frequentist
  properties of the method. We also illustrate their use in a
  motivating application to estimate the proportion of embolic strokes
  of undetermined source arising from occult cardiac sources or
  large-artery atherosclerotic lesions. Though we focus on the
  motivating example of the proportion of cases in an unlabeled
  dataset, the procedure is general and can be adapted to estimate any
  pathwise differentiable parameter in a non-parametric model.
\end{abstract}

\section{Introduction}\label{sec:intro}

The interpretation of the mathematical concept of probability is the
source of a historical divide of statistical methods between Bayesian
and frequentist \citep{fienberg2006did}. An \emph{objective}
interpretation of probability as the frequency of events is often
associated with frequentist statistics; a \emph{subjective}
interpretation as a representation of a state of knowledge or the
quantification or a subjective belief about nature is related to
Bayesian statistics
\citep{cox1946probability,de2017theory}. Additionally, Bayesian and
frequentist data analysis can be performed in \emph{(semi)-parametric}
or \emph{non-parametric} models, depending on the dimension assumed
for the parameters indexing the model. Nonparametric inference methods
are often preferred as they help avoid critical reliance on
assumptions on the functional form of the data probability
distributions, which are often scientifically unjustifiable. The
majority of the Bayesian statistics literature focuses on studying the
prior and posterior probabilistic behavior of the index of the
statistical model, whether the index is an Euclidean or
infinite-dimensional parameter. When the object of scientific inquiry
is characterized by a low-dimensional functional of the data
distribution, there is a dissonance between non-parametric Bayesian
methods, whieh require priors on full data distributions, and the
subjective interpretation of probability in terms of the state of
knowledge on the low-dimensional parameter.
This dissonance is partly responsible for the widespread use of
non-informative priors in Bayesian inference
\citep{kass1996formal}. In an attempt to remedy this issue, the task
of ``prior elicitation'', by which a prior on the parameters indexing
the model is constructed from available scientific knowledge, has
received some attention. However, most available methods work only for
parametric models \citep[e.g.,][]{chaloner1996elicitation,
  Ibrahim1998,chen1999prior,chen2003prior,albert2012combining}, and
methods for the nonparametric case are scarce. Among the few methods
in non-parametric Bayes, \cite{kessler2015marginally} propose a
solution by means of marginally specified priors, in which the prior
distribution is decomposed into two parts: an informative prior on a
finite set of functionals, and an informative prior on the rest of the
infinite-dimensional parameter. \cite{bush2010minimally} discuss the
elicitation of a Dirichlet process prior distribution in the context
of analysis of variance. Other methods attempt to specify the prior
trough empirical Bayes, i.e., estimating it from data
\citep[e.g.,][]{escobar1995bayesian,mcauliffe2006nonparametric}.

In this paper, we introduce a solution to the above problem through a
method called \emph{targeted Bayesian learning}, which is a method to
update the subjective belief on a low-dimensional functional of
interest representing the object of scientific inquiry. Prior and
posterior distributions are constructed to reflect prior and posterior
knowledge about specific target phenomena represented by
low-dimensional functionals, in contrast to non-parametric Bayesian
methods that focus on whole probability distributions. Similar to our
goal, \cite{bissiri2016general} (see also PAC Bayesian learning, e.g.,
\cite{mcallester1999some}) propose a procedure to update the belief
distribution of a parameter defined as the minimizer of a risk
function. We present more general methods that can be used to update
the belief distribution of a parameter defined as any pathwise
differentiable non-parametric functional (i.e., not only risk
minimizers) of the observed data distribution. Our methods are rooted
in the theory for efficient estimation of low-dimensional parameters
in general semi-parametric models, of which the foundational
frequentist concepts were laid by \cite{stein1956efficient,
  koshevnik1977non, pfanzagl1982contributions, begun1983information,
  vanderVaart91, newey1994asymptotic, vanderVaart&Wellner96,
  Bickel97}, among others. The theory is based on notions of
functional analysis and differential geometry. Of central importance
is the concept of a least-favorable submodel, loosely defined as any
parametric submodel that achieves the non-parametric efficiency bound
for the target parameter. This theory has led to a number of
frequentist estimation methods, for example, see
\cite{robins1994estimation, vanderLaan2003, Bang05, Tsiatis06,
  chernozhukov2018double} for methods based on estimating equations,
\cite{pfanzagl1982contributions} for one-step Newton-Raphson
corrections. Our proposed Bayesian method is more closely related the
targeted maximum likelihood estimation (targeted MLE) framework
\citep{vanderLaanRose11,vanderLaanRose18}, which can be loosely
described as computation of the MLE of the target parameter in (an
estimate of) the least-favorable parametric submodel. The proposed
targeted Bayesian learning proposal follows trivially from direct
analogy to classical Bayes: one only needs to specify the likelihood
in the parametric least-favorable submodel, and ``turn the Bayesian
crank'' to obtain a targeted posterior distribution of the parameter
in the (estimated) least favorable submodel. The posterior on the
target parameter is thus computed by applying the corresponding map to
the least favorable submodel. This surprisingly simple but powerful
idea was first proposed in a technical report \cite[][page
178]{vanderLaan08b} in the context of sequential adaptive
designs. \cite{Diaz2011} demonstrated its use in the context of
estimation of the average treatment effect of a binary exposure in an
observational study, but no further work on this area exists. The
least favorable submodel is often known up to a nuisance parameter
that must be estimated. We allow for estimation of these nuisance
parameters to be performed using flexible machine learning estimators.

Our approach also facilitates the asymptotic analysis of the posterior
distribution, compared to non-targeted non-parametric Bayesian
methods. Typical results for non-parametric Bayes include consistency
in the Hellinger distance, but obtaining convergence rates is usually
difficult and requires strong assumptions
\citep{Wasserman1998,freedman1999wald}. In contrast, targeted Bayesian
learning yields convergence of the posterior distribution at
$n^{1/2}$-rate to the asymptotic distribution of the targeted
MLE. This is in complete analogy to parametric Bayesian analysis in
which the posterior is shown to converge to the asymptotic
distribution of the MLE \citep[see the Bernstein-von Mises theorems
in][]{kleijn2012bernstein, bickel2012semiparametric}. Our results thus
show that the Bayesian posterior distribution inherits important
properties of the targeted MLE such as local efficiency and double
robustness.

Related to the above properties, several authors have proposed methods
to endow Bayesian methods for causal inference parameters with
frequentist properties such as double robustness and efficiency. These
methods proceed using one of three strategies: constructing a
pseudo-likelihood function where each observation is weighted using
inverse-probability weights \cite[e.g.,][]{saarela2015bayesian},
modifying the expected utility to include importance sampling weights
\cite[e.g.,][]{saarela2016bayesian}, or performing non-parametric
Bayes on the full distribution and computing the posterior of the
expectation of the doubly robust estimating equation
\cite[e.g.,][]{antonelli2018bayesian}. \cite{robins2015discussion}
critiqued these methods on the grounds that ``Bayesian logic is
rigidly defined: given a likelihood and a prior, one turns the
Bayesian crank to obtain a posterior. There is no wiggle room.'' The
main argument, based on the Robins-Ritov paradox
\citep{robins1997toward}, is that the target parameter is not a
function of the propensity score, and therefore the target posterior
cannot depend on the propensity score. Our method naturally
incorporates the propensity score through the Fisher information of
the least-favorable submodel, thus providing a natural way to perform
Bayesian non-parametric estimation with desirable frequentist
properties such as efficiency and double robustness.



We present the methods and ideas in the context of estimating the
proportion of cases in an unlabeled dataset. This choice of parameter
is motivated by an application to the study of the relation between
stroke and cardiac embolism. Embolic strokes of undetermined source
(ESUS) are thought to arise mostly from occult cardiac sources or
large-artery atherosclerotic lesions and, less frequently, other
causes. However, the proportion ESUS arising from occult cardiac
sources remains unclear. Knowing the proportion of cardioembolic
strokes is important because these patients may benefit from
anticoagulant therapy for secondary stroke prevention. Prior studies
indicate that this proportion could be as large as 30\%
\citep{montero2016asco}, but such estimates are based on imperfect
clinical inferences. The goal of our study is to estimate this
proportion based on the Cornell Acute Stroke Academic Registry
(CAESAR), which includes all patients (approx. 1700) with acute
ischemic stroke at our hospital from 2011-2016 and 187 features
extracted from echocardiography, while taking into account the
knowledge available from prior studies that the probability may be
around 30\%.  Though we present the methods in the context of this
motivating example, the ideas developed here are of general
applicability to any pathwise differentiable parameter in the
non-parametric model.

\section{Inferential problem}\label{sec:problem}

We let $X$ denote a vector of features or predictors (e.g., features
of the echocardiogram), let $Y\in\{0,1\}$ denote a binary outcome
(e.g., whether a stroke is of cardiac source), and let $L\in\{0,1\}$
denote a binary variable equal to one if the data corresponding to $X$
is labeled (e.g., whether the source of the stroke is known), and $0$
otherwise. We let $\Q$ denote the distribution of the triplet
$Z=(X,L,LY)$ in the population of interest, and assume that $\Q$ is
dominated by a measure $\nu$ with density $\q$. We let $\q$ be an
element of the non-parametric model $\M$. For a function $f:z\to \R$
we use $\Q f$ to denote $\int f\dd\Q$. In this paper we focus on
estimating $\theta =\E(Y\mid L=0)$, where $\E(\cdot)$ denotes
expectation with respect to $\Q$. The parameter $\theta$ is often
referred to as the class proportion in unlabeled data.

The methods we present are valid under random and matched-cohort
sampling from $\Q$. In matched-cohort studies, a sample of $n_0$ of
unlabeled units is observed along with $n_1=kn_0$ labeled units, where
$k$ labeled units are sampled for each unlabeled unit. Unlabeled and
labeled units can be matched on a subset of discrete features
$W\subseteq X$. We use $\bar W=X\,\backslash\, W$ to denote the
features that are not used for matching. Under no matching we have
$W=\emptyset$ and we simply observed two separate samples from labeled
and unlabeled units. In the sequel we assume $Z_1,\ldots,Z_n$ denotes
an i.i.d. sample from a distribution $\P$, where $\P=\Q$ for random
sampling and
\[\p(Z) = \q(LY\mid L, X) \q(\bar W\mid W, L)\q(W\mid L=0)\r^L(1-\r)^{1-L}\]
for matched-cohort sampling \citep{kennedy2015semiparametric}. We will
sometimes use the notation $\bar Z_n=(Z_1,\ldots,Z_n)$. Here
$\r=k/(k+1)$ is the probability that $L=1$ under $\P$. Under random
sampling we define $\r=\P(L=1)$.

Let $\Pn$ denote the empirical distribution of $Z_1,\ldots,Z_n$, let
$\m(x)=\EQ(Y\mid L=1, X=x)$ denote the outcome expectation (class
probability) in the labeled data, and $\g(x)=\P(L=1\mid X=x)$ denote
the sampling mechanism that gives rise to the biased sample of labeled
data. Note that $\m(x)$ is the same under random and matched-cohort
sampling, but $\g(x)$ may differ. Clearly, $\theta =\E(Y\mid L=0)$ is
not identifiable from $\P$ without additional assumptions. Throughout
the paper we make the following assumptions:
\begin{assumptioniden}[Exchangeability]
  Assume $\E(Y\mid L=1, X)=\E(Y\mid L=0, X)$.\label{ass:random}
\end{assumptioniden}
\begin{assumptioniden}[Overlap]
Assume $\P\{\g(X)>\epsilon\mid L=0\}=1$ for some $\epsilon >0$.\label{ass:overlap}
\end{assumptioniden}
Assumption \ref{ass:random} is standard in missing data problems
\citep{Rubin74} and is satisfied if units are labeled/unlabeled
completely at random within strata of the features $X$. Assumption
\ref{ass:overlap} states that there is enough overlap between the $X$
distributions of labeled and unlabeled units to allow
identification. Under these assumptions, the class probability
$\theta$ may be expressed as
\begin{align*}
  \theta &=\E\{\E(Y\mid L=0, X)\mid L=0\}=\EQ\{\m(X)\mid L=0\},
\end{align*}
where the first equality follows from the rule of iterated expectation
and the second from \ref{ass:random} and the definition of
$\m$. Assumption \ref{ass:overlap} is necessary so that conditional
expectations are supported in the data, and to ensure regularity of
the parameter.

We will sometimes use $\Theta$ to denote the parameter functional
$\P\mapsto \EQ\{\m(X)\mid L=0\}$ that maps any distribution $\P\in\M$
into a parameter value in $[0,1]$. In \S\ref{sec:frequentist}
below we discuss non-parametric efficient estimation of $\theta$ in a
frequentist setting, reviewing relevant concepts such as the
efficiency bound and the least favorable parametric sub-model. These
concepts will then serve as the foundation for the proposal of
Bayesian non-parametric estimators developed in
\S\ref{sec:bayesian}, which constitute the main contribution of
this manuscript.

\section{Non-parametric efficient frequentist
  estimation}\label{sec:frequentist}
We start this section by reviewing two fundamental and related
concepts in efficient non-parametric estimation theory: the efficient
influence function (EIF) and least-favorable submodel. These
concepts are central to the Bayesian procedure proposed in
\S\ref{sec:bayesian}. We will provide a heuristic discussion, a
rigorous treatment may be found, for example, in \cite{Bickel97}. We
then review the targeted MLE for this parameter, originally proposed
by \cite{Hubbard2011}.

\subsection{Preliminaries in frequentist efficiency theory}
Consider a parametric submodel
$\M_\varepsilon=\{\p_\varepsilon:\varepsilon \in \R\}\subset \M$ such
that $\p_0=\p$. Estimating the parameter $\Theta(\P)$ in the model
$\M$ is clearly more difficult than estimating it in $\M_\varepsilon$
\citep{vanderVaart98}. The efficiency bound in $\M$ can thus be
defined as the supremum of the Cram\'er-Rao bounds of all parametric
submodels $\M_\varepsilon$ that locally cover $\M$. The least
favorable submodel, also referred to as the ``hardest'' submodel, can
thus be defined as any submodel achieving this efficiency bound.

More precisely, for a mean-zero function $h$ such that
$||h||_{\infty}<\infty$, define a submodel
$\{\p_{\varepsilon,h}:\varepsilon\}$ such as the exponential tilting
model $\p_{\varepsilon,h}(z) \propto \exp\{\varepsilon
h(z)\}\p(z)$. Note that the score of this model is precisely the
function $h$, and that $h$ uniquely determines the submodel and
therefore there is a correspondence between submodels and score
functions. The Cram\'er-Rao bound for estimating $\theta$ in
$\M_\varepsilon$ is given by
\begin{equation}
  \frac{\dd}{\dd \varepsilon}\Theta(\P_{\varepsilon,h})\bigg|_{\varepsilon =
    0}\frac{1}{\P h^2}.\label{eq:crbound}
\end{equation}
This expression reveals that efficient semiparametric inference is
only possible for parameters that are smooth enough in the sense that
they are \textit{pathwise differentiable}, i.e., in the sense that the
derivative in the above expression exists. In that case, the Riesz
representation theorem shows that the derivative must admit the
representation
\begin{equation}
  \frac{\dd}{\dd \varepsilon}\Theta(\P_{\varepsilon,h})\bigg|_{\varepsilon =
    0}=\P\lambda h,\label{eq:pwdiff}
\end{equation} for some function $\lambda$ with $\P \lambda=0$ and
$\P \lambda^2<\infty$. Any such function $\lambda$ is called a
\textit{gradient} of the pathwise derivative. Importantly, this gradient is not uniquely
defined for general semi-parametric models, but it is unique for
non-parametric models. Taking the supremum
of all Cram\'er-Rao bounds across score functions $h$, together with
pathwise differentiability of $\Theta$, the Cauchy-Schwarz inequality yields
\[\sup_h\frac{\P \lambda h}{\P h^2}\leq \P \lambda^2,\]
with equality only if $\lambda$ is an element of the closed linear
span of the score space $L_2^0(\P)$ of all functions $h$ with mean
zero and finite variance under $\P$. The unique projection of any
gradient onto $L_2^0(\P)$ is referred to as the \textit{canonical
  gradient}, \textit{efficient influence function} (EIF), or
\textit{efficient score}. In the sequel we use $\lambda$ to denote the
EIF in a slight abuse of notation. This function characterizes the
semiparametric efficiency bound. In particular, all regular estimators
of $\theta$ have an asymptotic variance larger than or equal to
$\P\lambda^2$ \cite[see the H\'ajek convolution theorem
in][]{hajek1970characterization,Bickel97}. The convolutiuon theorem in
fact proves a stronger result: it states that the optimal asymptotic
distribution for any estimator $\hat\theta$ is
$\sqrt{n}(\hat\theta-\theta)\rightsquigarrow N(0,\P\lambda^2)$, where
we use $\rightsquigarrow$ to denote weak convergence.

For the particular case of the parameter $\Theta(\P)=\E\{\m(X)\mid L=0\}$,
the efficient influence function under random sampling is given by
\citep{hahn1998role}:
\[\lambda_\eta(Z)=\frac{L}{\g(X)}\frac{1-\g(X)}{1-\r}\{Y-\m (X)\} +
  \frac{1-L}{1-\r}\{\m(X) - \theta\},\] where we have added the index
$\eta=(\m,\g)$ to the notation to highlight the dependence of $\lambda$
on these nuisance parameters. \cite{kennedy2015semiparametric} showed that the same influence
function is valid under matched-cohort sampling. The efficiency bound
is thus given by $\var\{\lambda_\eta(Z)\} = \P\lambda_\eta^2=\P\psi_\eta$, where we define
\begin{equation}
  \psi_\eta(z) =
  \frac{1-\g(x)}{(1-\r)^2}\left\{\frac{1-\g(x)}{\g(x)}\sigma^2(x) +
    \{\m(x)-\theta\}^2\right\},\label{eq:psi}
\end{equation}
and $\sigma^2(x) =\var(Y\mid L=1,X=x)$. We will now proceed to
construct a parametric submodel that attains this efficiency
bound. The frequentist targeted maximum likelihood estimator reviewed
in \S\ref{sec:tmle} proceeds by performing MLE in this submodel, by means
of an iterated procedure that updates initial estimators
$\hat\eta=(\hat\m,\hat\g)$ until the score equations of the submodel
are solved. This submodel is also at the center of our Bayesian
targeted proposal, as it provides an appropriate likelihood to use for
updating a prior distribution on $\theta$ that guarantees desirable
properties of the posterior distribution such as double robustness and
efficiency.

\subsection{Least favorable submodel}
The construction of a least favorable submodel will be based on the
factorization of the likelihood
$\p(Z) = \p(Y\mid L,X)\p(L\mid X)\p(X)$.  The first step is to
decompose the efficient score $\lambda_\eta(z)$ into scores for each
of the conditional probabilities in this factorization. This
decomposition may be achieved by projecting $\lambda_\eta(Z)$ into the
space of scores for each model. For example, the score in the model
for $\p(Y\mid L,X)$ is obtained by taking the projection of
$\lambda_\eta(Z)$ onto the space of functions of $Z$ with conditional
expectation given $(L,X)$ equal to zero. This process yields the
following scores:
\begin{alignat*}{3}
  \lambda_{Y,\eta}(Z) &= H(X)L\{Y-\m
                     (X)\},\,&&\text{ where }\, H(x)=&&\frac{1}{\g(x)}\frac{1-\g(x)}{1-\r};\\
  \lambda_{L,\eta}(Z) &= M(X)\{L-\g (X)\},\,&&\text{ where }\, M(x)=-&&\frac{\m(x) -\theta}{1-\r};\\
  \lambda_{X,\eta}(X) &= \frac{1-\g(X)}{1-\r}\{\m(X) -\theta\}.&&&&
\end{alignat*}
We can now use these scores to construct exponential tilting least
favorable submodels. For a univariate parameter
$\varepsilon\in\R$, we use the following submodels
\begin{align}
  \logit \m_\varepsilon(x) &= \logit \m(x) + \varepsilon H(x),\notag\\
  \logit \g_\varepsilon(x) &= \logit \g(x) + \varepsilon M(x),\label{eq:submodel}\\
  \p_\varepsilon(x) &\propto \exp\{\varepsilon \lambda_{X,\eta}(x)\}\p(x)\notag,
\end{align}
where $\logit(p)=\log\{p/(1-p)\}$. It is easy to see that the scores
of these submodels span the score $\lambda_\eta(Z)$, and that these
submodels are such that $\p_0=\p$. 
We will sometimes use $\hat\m_\varepsilon(x)$,
$\hat\g_\varepsilon(x)$, and $\hat\p_\varepsilon(x)$ to denote the
above submodels with the true quantities $\m(x)$, $\g(x)$, and $\p(x)$
replaced by initial estimates $\hat\m(x)$, $\hat\g(x)$, and
$\hat\p(x)$.

\subsection{Cross-fitted targeted maximum likelihood
  estimation}\label{sec:tmle}
In this section we briefly discuss the construction of the targeted
MLE for $\theta$. We first discuss the classical (that is, not
cross-fitted) version of the targeted MLE, and then discuss its
cross-fitted version that may be used to incorporate flexible
data-adaptive methods in estimation of $\eta$. The interested reader
is encouraged to consult \cite{Hubbard2011} for a more complete
treatment of the targeted MLE for $\theta$, \cite{zheng2011cross} for
cross-fitting in targeted MLE, and \cite{vanderLaanRose11,vanderLaanRose18}
for the targeted MLE of a variety of parameters.

Computation of the targeted MLE of $\theta$ proceeds by estimating the
parameter $\varepsilon$ in the submodel (\ref{eq:submodel}),
proceeding in an iterative fashion: (i) start with initial estimators
$\hat\eta$, and estimate the parameters in $\hat\m_\varepsilon(x)$,
$\hat\g_\varepsilon(x)$, and $\hat\p_\varepsilon(x)$ with MLE
(treating $\hat\eta$ as fixed), and (ii) update
$\hat\m(x)=\hat\m_{\hat\varepsilon}(x)$,
$\hat\g(x)=\hat\g_{\hat\varepsilon}(x)$, and
$\hat\p(x)=\hat\p_{\hat\varepsilon}(x)$. The iteration is repeated
until convergence is achieved, that is, until
$\hat\varepsilon = o_P(n^{-1/2})$. This implies
$n^{-1}\sum_{i=1}^n\lambda_{\tilde\eta}(Z_i)=o_P(n^{-1/2})$, where we
use $\tilde\m (x)$, $\tilde\g(x)$, and $\tilde\p(x)$ to denote the
last estimates in the iteration, i.e., the maximum likelihood
estimates targeted towards estimation of $\theta$. The classical
targeted MLE of $\theta$ is thus defined as
\[\hat\theta = \frac{1}{1-\hat\r}\frac{1}{n}\sum_{i=1}^n\{1-\tilde\g(X_i)\}\tilde \m(X_i).\]
For the frequentist targeted MLE, \cite{Hubbard2011} use three
different parameters in each of the submodels (\ref{eq:submodel}). In
the frequentist case, this choice is more computationally
convenient. In particular, when the empirical distribution $\P_n(x)$
is used to estimate $\P(x)$, and the submodels are built with three
different parameters, it is not necessary to fluctuate $\P_n(x)$ as it
solves all score equations by definition. In addition, the fluctuation
of the initial estimators for $\m(x)$ and $\g(x)$ may be carried out
using existing software and methods for logistic regression. For the
purpose of Bayesian estimation, the choice of a single parameter
$\varepsilon$ in the submodels (\ref{eq:submodel}) will prove more
convenient, as discussed below.

We introduce the following key assumption on the initial estimators of
the nuisance parameters:
\begin{assumptioniden}[Consistency of second order
  term] \label{ass:rate} Assume
  $||\hat\g - \g|| ||\hat\m - \m|| = o_P(n^{-1/2})$.
\end{assumptioniden}
This assumption may be satisfied if both $\m(x)$ and $\g(x)$ are
consistently estimated at certain slow (e.g., $n^{1/4}$) rates. To
increase assurance that this assumption is satisfied, targeted MLE
departs from the classical but unrealistic parametric setting by
allowing usage of flexible data-adaptive estimators from the machine
and statistical learning literature. The required rates are achievable
by many data-adaptive regression algorithms. See for example
\cite{bickel2009simultaneous} for rate results on $\ell_1$
regularization, \cite{wager2015adaptive} for rate results on
regression trees, and \cite{chen1999improved} for neural networks. The
assumption may also be satisfied by the highly adaptive lasso
\citep[HAL,][]{benkeser2016highly} under the condition that the true
regression functions are right-hand continuous with left-hand limits
and have variation norm bounded by a constant. Ensemble learners such
as the super learner \citep{vanderLaanPolleyHubbard07} may also be
used. Super learning builds a combination of predictors in a
user-given library of candidate estimators, where the weights minimize
the cross-validated risk of the resulting combination, and has
important theoretical guarantees
\citep{vanderLaanDudoitvanderVaart06,vanderVaartDudoitvanderLaan06}
such as asymptotic equivalence to the oracle selector.

The analysis of classical targeted MLE relies on the powerful but
restrictive empirical processes Donsker condition that the estimators
$\hat\g$ and $\hat\m$ fall in function classes with bounded
complexity, specifically bounded entropy integrals. Donsker conditions
may be inappropriate when data-adaptive estimators, which live in
complex spaces, are allowed for $\g$ and $\m$. For example functions
classes with unbounded variation are generally not Donsker, and highly
adaptive estimators such as random forests may have unbounded
variation. Fortunately, complexity assumptions may be avoided by
introducing cross-fitting into the estimation procedure. Cross-fitting
was first used in the context of targeted minimum loss-based
estimation by \cite{zheng2011cross}, and has also been applied to
Neyman-orthogonal estimating equations
\citep{chernozhukov2016double}. Let ${\cal V}_1,\ldots,{\cal V}_J$
denote a random partition of the index set $\{1,\ldots,n\}$ into $J$
validation sets of approximately the same size. That is,
${\cal V}_j\subset \{1,\ldots,n\}$;
$\bigcup_{j=1}^J {\cal V}_j = \{1,\ldots,n\}$; and
${\cal V}_j\cap {\cal V}_{j'}=\emptyset$. In addition, for each $j$,
the associated training sample is given by
${\cal T}_j=\{1,\ldots,n\}\setminus {\cal V}_j$. Denote by
$\hat \eta_{{\cal T}_j}$ the estimator of $\eta$ obtained by training
the corresponding prediction algorithms using only data in the sample
${\cal T}_j$. Let also $j(i)$ denote the index of the validation set
which contains observation $i$. The cross-fitted targeted MLE
$\tilde \eta_{{\cal T}_{j(i)}}(X_i)$ is constructed replacing
$\hat \eta(X_i)$ by its cross-fitted version
$\hat \eta_{{\cal T}_{j(i)}}(X_i)$ as the initial estimator in the
TMLE algorithm.  The cross-fitted targeted MLE of $\theta$ is then
given by
\[\tilde\theta = \frac{1}{1-\hat\r}\frac{1}{n}\sum_{i=1}^n\{1-\tilde\g_{{\cal T}_{j(i)}}(X_i)\}\tilde \m_{{\cal T}_{j(i)}}(X_i).\]
For posterior reference, we present the following result due to
\cite{van2010estimation} and \cite{zheng2011cross}, which establishes
the asymptotic optimality of the cross-fitted targeted MLE.

\begin{theorem}[Asymptotic linearity of cross-fitted targeted
  MLE]\label{theo:aslin} Assume \ref{ass:rate}.
  Then the cross-fitted targeted MLE is asymptotically linear and
  efficient:
\[\sqrt{n}(\tilde\theta-\theta)=\frac{1}{\sqrt{n}}\sum_{i=1}^n\lambda_\eta(Z_i)
  + o_P(1).\]
\end{theorem}

\section{Targeted non-parametric Bayesian
  estimation}\label{sec:bayesian}

We now turn to describing our proposal to obtain a posterior
distribution for $\theta$ based on a prior distribution and an
i.i.d. sample $Z_1,\ldots,Z_n$. The procedure is carried out in three
steps. First, we map a prior distribution $\Pi_\theta$ on $\theta$ to
a prior distribution on the parameter $\varepsilon$ of the least
favorable submodel. Second, we estimate the posterior distribution on
$\varepsilon$ using the likelihood of the (cross-fitted) least
favorable submodel. And lastly, we map the posterior on $\varepsilon$
back into a posterior on $\theta$. We then prove that this algorithm
results in a posterior distribution which converges to the
distribution of the cross-fitted targeted MLE. The details of this
process are described below.

In the sequel, for notational convenience we let $\tilde \eta(X_i)$
denote the targeted cross-fitted MLE
$\tilde \eta_{{\cal T}_{j(i)}}(X_i)$.

\paragraph{Mapping a prior on $\theta$ to a prior on $\varepsilon$.}
The prior distribution $\Pi_\theta(\theta)$ is assumed to have a
density $\pi_\theta(\theta)$ with respect to the Lebesgue
measure. This density may be mapped into a prior density on
$\varepsilon$ using the map
\begin{equation}
\vartheta:\varepsilon \mapsto \Theta(\P_\varepsilon)=\frac{1}{1-\r}\int\{1-\g_\varepsilon(x)\}
  \m_\varepsilon(x)\p_\varepsilon(x)\dd\nu(x).\label{eq:map}
\end{equation} Denoting
$\vartheta'(\varepsilon)$ the derivative of this map, a prior density for $\varepsilon$ may be
constructed as
\[\pi_\varepsilon(\varepsilon) = |\vartheta'(\varepsilon)|\,
  \pi_\theta[\vartheta(\varepsilon)].\] We let
$\tilde\pi_\varepsilon(\varepsilon)$, $\tilde\vartheta(\varepsilon)$,
and $\tilde\vartheta'(\varepsilon)$ denote the corresponding
quantities computed when the unknown parameters $\m(x)$, $\g(x)$, and
$\p(x)$ in $\vartheta(\varepsilon)$ and $\vartheta'(\varepsilon)$ are
replaced by the cross-fitted targeted maximum likelihood estimates
$\tilde\m(x)$, $\tilde\g(x)$, and $\tilde\p(x)$ obtained in the last
step of the iterative cross-fitted targeted MLE procedure.

\paragraph{Posterior distribution.} Given the prior density
$\tilde\pi_\varepsilon(\varepsilon)$, and the likelihood of the least
favorable submodel (\ref{eq:submodel}), the posterior density on
$\varepsilon$ is trivially given by
\[\tilde\pi_\varepsilon(\varepsilon\mid \bar Z_n)\propto
  \tilde\pi_\varepsilon(\varepsilon)\tilde\p(\bar Z_n\mid
  \varepsilon),\] where
\begin{equation}
\tilde\p(\bar Z_n\mid \varepsilon)=\prod_{i=1}^n
  \left[\tilde\m_\varepsilon(X_i)^{Y_i}\{1-\tilde\m_\varepsilon(X_i)\}^{1-Y_i}\right]^{L_i}\tilde\g_\varepsilon(X_i)^{L_i}\{1-\tilde\g_\varepsilon(X_i)\}^{1-L_i}\tilde\p_\varepsilon(X_i)\label{eq:likeli}
\end{equation}
is the likelihood in the least favorable submodel.  Then, for a set
$\mathbb B\subseteq [0,1]$, its posterior measure with respect to
$\theta$ is simply the measure of the preimage of $\mathbb B$ under
$\tilde\vartheta$ with respect to $\varepsilon$. That is,
\[\tilde\Pi_\theta(\mathbb B\mid \bar Z_n) =
  \tilde\Pi_\varepsilon\{\tilde\vartheta^{-1}(\mathbb B)\mid \bar
  Z_n\},\] where
$\tilde\Pi_\varepsilon(\mathbb A\mid \bar Z_n)=\int_{\mathbb
  A}\tilde\pi_\varepsilon(\varepsilon\mid \bar Z_n)\dd\varepsilon$
denotes the posterior measure of $\mathbb A$ with respect to
$\varepsilon$, and
$\tilde\vartheta^{-1}(\mathbb B)=\{\varepsilon \in
\R:\tilde\vartheta(\epsilon) \in \mathbb B\}$. A numerical
approximation to this posterior distribution of $\theta$ may be
obtained by sampling a large number $K$ of observations
$\varepsilon_k:k\in\{1,\ldots,K\}$ from the density
$\tilde\pi_\varepsilon(\varepsilon\mid \bar Z_n)$, and then evaluating
$\theta_k=\tilde \vartheta(\varepsilon_k)$. Empirical quantiles and
moments of the sample $\theta_k:k\in\{1,\ldots,K\}$ may be used to
approximate quantiles and moments of
$\tilde\Pi_\theta(\cdot\mid \bar Z_n)$, where this approximation can
be made arbitrarily accurate by letting $K\to\infty$, up to
computational restrictions.

In some applications it may be desirable to work directly with the
posterior density of $\theta$, rather than its posterior
distribution. The analytical expression of that density may be easily
found under the following assumption on the map $\vartheta$:
\begin{assumptioniden}[Piecewise smooth invertibility of $\vartheta$]
  Assume that $\R$ may be partitioned into (possibly countably
  infinite) intervals $\I_j:j\in\{1,\ldots,\infty\}$ such that
  $\vartheta(\varepsilon)$ is equal to some $\vartheta_j(\varepsilon)$
  in $\I_j$ and $\vartheta_j$ has inverse function $\chi_j$ with
  derivative $\chi'_j$.\label{ass:pwsi}
\end{assumptioniden}
Under \ref{ass:pwsi}, the posterior density of $\theta$ may be written
as
\begin{align*}
  \tilde\pi_\theta(\theta\mid \bar Z_n) &= \sum_{j=1}^\infty
                                        \one_{\I_j}\{\tilde\chi_j(\theta)\}\tilde
                                        \pi_\varepsilon\{\tilde\chi_j(\theta)\}\tilde{\chi}'_j(\theta)\\
                                      &\propto \pi_\theta(\theta)\sum_{j=1}^\infty
                                        \one_{\I_j}\{\tilde\chi_j(\theta)\}
                                        \tilde\p\{\bar Z_n\mid\tilde\chi_j(\theta)\}.
\end{align*}
As expected, it is easy to see that this posterior density corresponds
to the classical Bayesian posterior obtained through a
reparameterization of the likelihood of the least favorable submodel in terms of
$\theta=\tilde\vartheta(\varepsilon)$. 

\subsection{Bernstein von-Mises type asymptotic convergence}

In parametric Bayesian inference, the marginal posterior for the
parameter of interest is expected to be asymptotically Gaussian and
satisfy frequentist criteria for optimality such as efficiency. This
is summarized in a result known as the Bernstein von-Mises theorem
\citep{le2012asymptotic}. This result has been generalized to a number
of cases, including the parametric component of a model indexed by an
Euclidean and a finite-dimensional parameter
\citep{kleijn2012bernstein} and misspecified parametric models
\citep{bickel2012semiparametric}. The following theorem shows that an
analogous result holds for our targeted Bayesian proposal.

\begin{theorem}[Bernstein von-Mises]\label{theo:ber} Let $\bar\theta$
  denote a random variable distributed as $\tilde\Pi_\theta$. Assume:
  \begin{enumerate}[label=(\roman*)]
  \item The prior density $\pi_\theta$ is continuous and strictly positive at
    $\theta$.
  \item The targeted maximum likelihood estimator $\tilde\eta$ is such
    that $\psi_{\tilde\eta}$ is Glivenko-Cantelli with
    $\P\psi_{\tilde\eta} = \P\psi_\eta + o_\P(1)$.
  \end{enumerate}
Then the posterior distribution converges in total variation:
\[\sup_\B\big|\tilde\Pi_\theta\{\sqrt{n}(\bar\theta - \theta)\in \B\mid \bar Z_n\}
  - N_{\sqrt{n}(\tilde\theta - \theta),\P\psi_\eta}(\B)\big|\to 0,\]
where $\tilde\theta$ is the targeted MLE and $N_{\mu,\sigma^2}$ is the
Gaussian distribution with mean zero $\mu$ and variance $\sigma^2$.
\end{theorem}

This theorem, proved in the supplementary materials, shows that the
posterior distribution converges to a Gaussian variable centered at
the targeted MLE. As a consequence, and under the conditions of
Theorem~\ref{theo:aslin}, credible intervals based on
$\tilde\Pi_\theta(\cdot\mid \bar Z_n)$ have correct frequentist
coverage asymptotically. In other words, the theorem implies that
$(1-\alpha)100\%$ credible intervals based on quantiles converge to
$\tilde\theta \pm q_{\alpha/2}\sqrt{\P\psi_\eta/n}$, where
$q_{\alpha}$ is the $\alpha$ quantile of a standard normal
distribution. Under the conditions of Theorem~\ref{theo:aslin}, the
coverage of the latter interval is the nominal level
$(1-\alpha)100\%$.  Lastly, because the targeted MLE is doubly robust,
the posterior mean is also $n^{1/2}$-consistent for $\theta$ when both
$\tilde\g$ and $\tilde\m$ are consistent as in
Theorem~\ref{theo:aslin}. Centrality measures such as the posterior
mean, median, or mode are consistent whenever at least one of
$\tilde\g$ or $\tilde\m$ is consistent. In addition, the theorem shows
that the and the variance of the posterior distribution converges to
the efficiency bound $\P\psi_\eta$.


\section{Motivating application}

In order to estimate the proportion of ESUS arising from occult
cardiac sources, we leveraged data from the Cornell Acute Stroke
Academic Registry (CAESAR), containing 1663 patients, 1083 of which
had a cause of stroke adjudicated by a neurologist, and the remaining
of which are ESUS. In addition to 186 features extracted from the
echocardiography, our predictors $X$ included clinical and demographic
data such as age, sex, race, smoking status, previous stroke, type of
insurance, atrial fibrilation, chronic kidney disease, chronic heart
failure, depression, etc. Our label variable $Y$ is equal to one if
the adjudicated stroke is of cardiac sources, zero if not, and missing
if the stroke is ESUS. The prior information was encoded in a
$\text{Beta}(\alpha = 2.7, \beta=6.3)$ distribution. It is easy to
check that this distribution has a mean of $0.3$, corresponding to the
the proportion of cardioembolic strokes previously reported in the
literature \citep{montero2016asco}, while its standard deviation
approximately $0.16$, reflecting the fact that we are not very certain
about the mean value $0.3$.

\begin{figure}[H]
  \centerline{\includegraphics[width=4in]{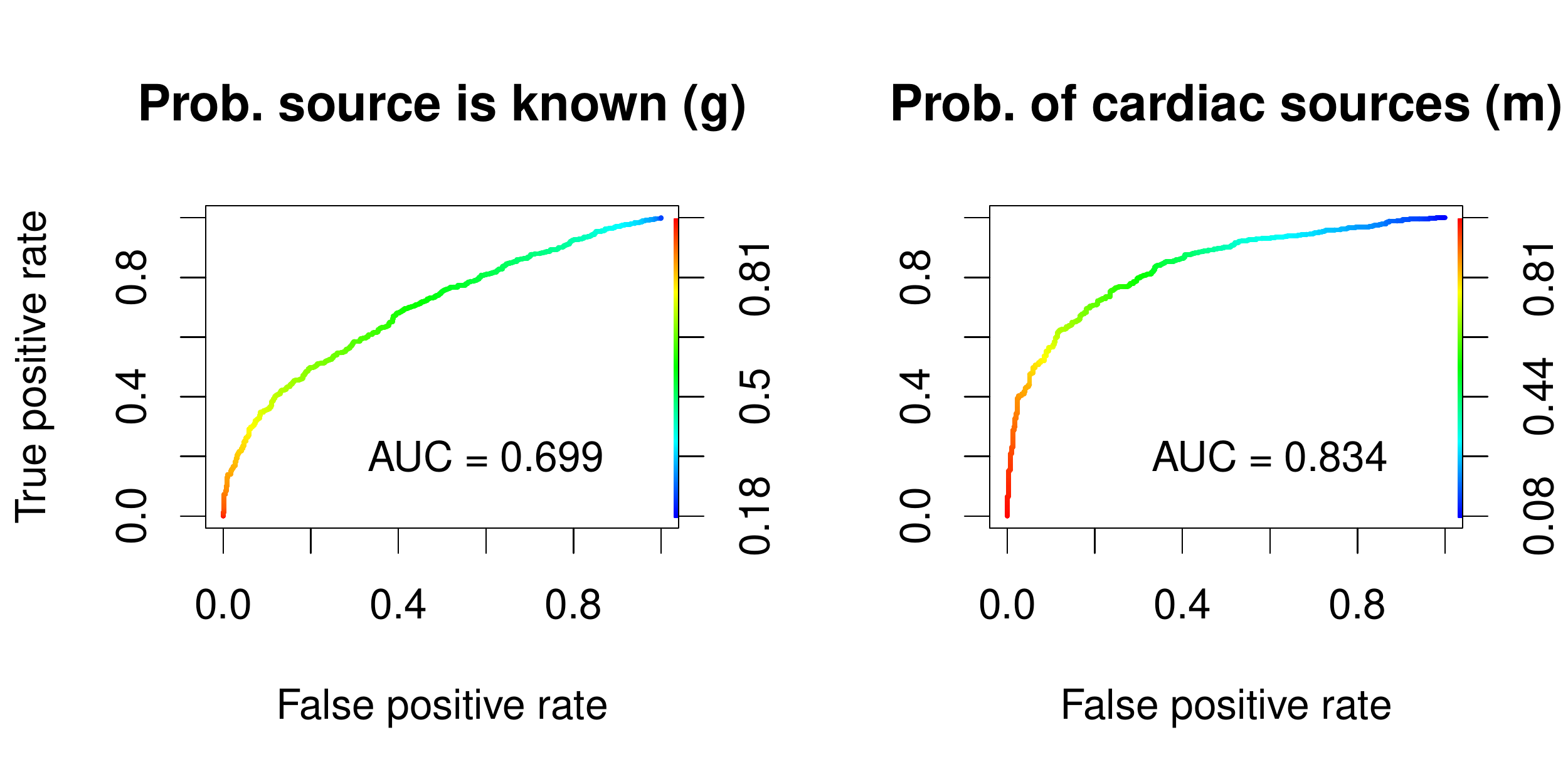}}
  \caption{ROC curve and area under the curve for cross-fitted predicted
    probabilities $\g(x)$ and $\m(x)$.}\label{fig:roc}
\end{figure}

In order to estimate the probability functions $\g(x)$ and $\m(x)$, we
used an ensemble known as the super learner
\citep{vanderLaanPolleyHubbard07,SL}, which builds a convex
combination of predictors in a user-supplied library, with weights
chosen to optimize the cross-validated prediction error. Our library
consisted of: random forests, extreme gradient boosting, multivariate
adaptive splines (MARS), logistic regression with $\ell_1$
regularization, and simple logistic regression with a pre-screening
algorithm that selects the 50 features with a larger correlation with
the outcome. Hyper-parameter tuning for the random forest and extreme
gradient boosting algorithms was performed using the \texttt{caret R}
package \citep{caret}. The \texttt{glmnet} package \citep{glmnet} was
used for regularized logistic regression, whereas MARS was computed
using the \texttt{earth} package \citep{earth}. In order to avoid the
Donsker conditions in Theorem~\ref{theo:aslin}, we cross-fitted the
super learner. Figure~\ref{fig:roc} shows the cross-fitted ROC curves
for each parameter.

\begin{figure}[H]
  \centerline{\includegraphics[width=2.2in]{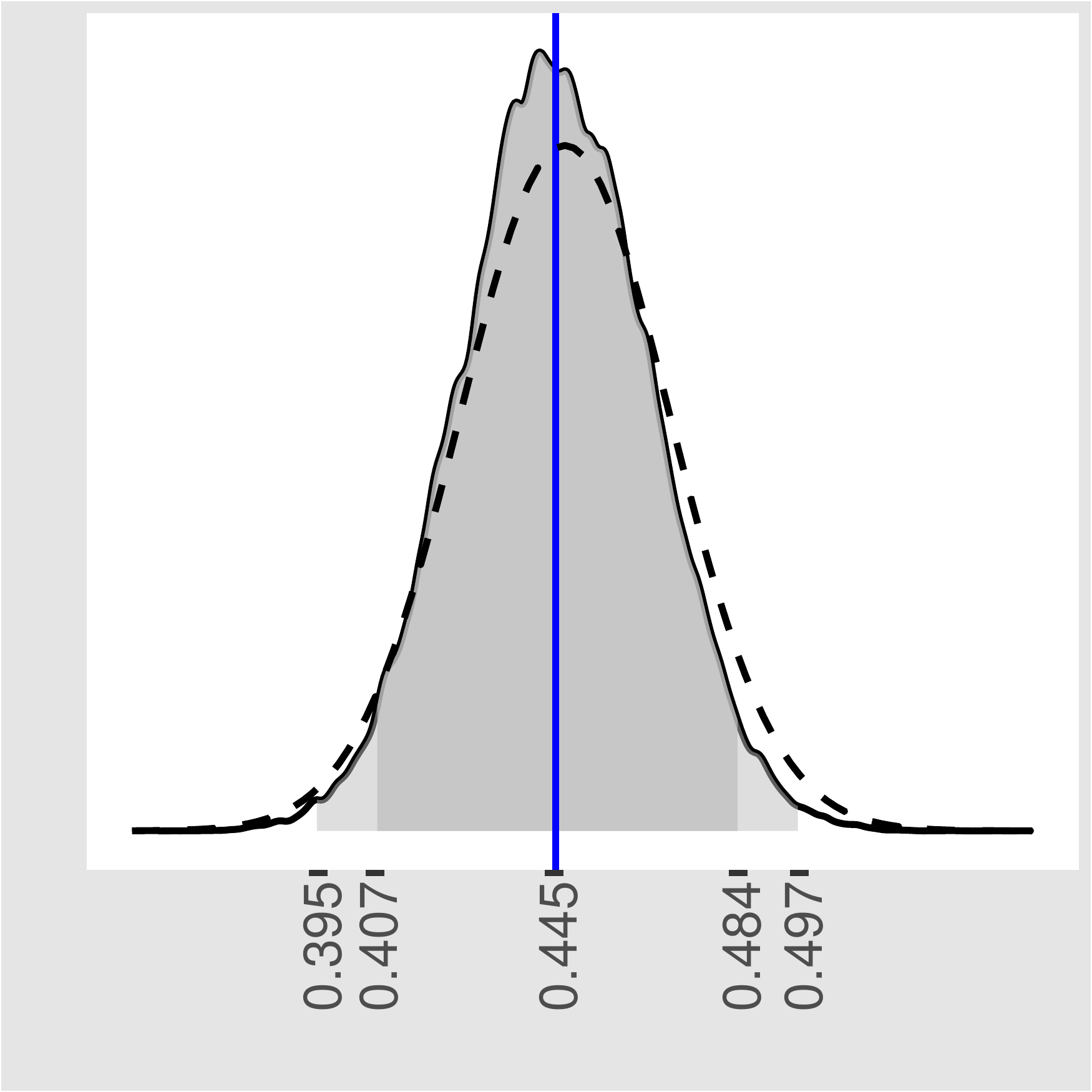}}
  \caption{Posterior distribution with 99\% (light gray) and 95\%
    (dark gray) credible intervals. The dashed line is a normal
    density with mean the targeted MLE and variance equal to the
    estimated variance of the EIF.}\label{fig:post_ci}
\end{figure}

Once the cross-fitted probabilities
$\hat \eta_{{\cal T}_{j(i)}}(X_i) = (\hat \g_{{\cal T}_{j(i)}}(X_i),
\hat \m_{{\cal T}_{j(i)}}(X_i))$ were computed with super learner, we
proceeded with our targeted learning Bayesian algorithm. We sampled
$10^6$ values from this posterior using the Metropolis Hastings
algorithm implemented in the library \texttt{mcmc} \citep{mcmc}. The
resulting posterior, along with 95\% and 99\% credible intervals, is
presented in Figure~\ref{fig:post_ci}, together with the normal
distribution centered at the targeted MLE $\tilde\theta$, and with
estimated variance given by the empirical variance of the EIF
$\lambda_{\tilde\eta}(Z_i)$. The proximity between this normal density
and the posterior density is an illustration of our Bernstein-von
Mises-type result in Theorem~\ref{theo:ber}.

This analysis allows us to conclude that the proportion of cardiogenic
ESUS is between $34.9\%$ and $45.2\%$ with $99\%$ probability. This
proportion is much higher than previous studies have suggested, and
supports the hypothesis that a substantial proportion of ESUS patients
may benefit from anticoagulant therapy for secondary stroke
prevention, but also underlines that the majority of ESUS cases are
not cardioembolic, which might explain the failure of previous
clinical trials of anticoagulant therapy in the overall ESUS
population \citep{hart2016rivaroxaban, diener2019dabigatran}.

\section{Numerical Studies}

We present the results of numerical studies aimed to evaluate the
performance of the proposed methods. We generated 1000 datasets for
each sample size $n \in \{ 400, 900, 1600, 2500, 4900\}$, using the
following data generating mechanism:
\begin{align*}
  X_j &\sim \text{Uniform}(-1, 1) \text{ for } j \in \{1,2,3\}\\
  L&\sim \text{Bernoulli}\{\g(X)\}\\
  Y\mid L = 1&\sim \text{Bernoulli}\{\m(X)\},
\end{align*}
where the probability functions are given by
\[ \g(X) = \text{expit}\left(\sum_{j = 1}^{3}X_i \right)\text{ and }
  \m(X) = \text{expit}\left(1 - \sum_{j = 1}^{3}X_i \right).\] We
consider four different scenarios for estimation of the nuisance
parameters: (a) both $\g$ and $\m$ are consistently estimated, (b)
only $\g$ is consistently estimated, (c) only $\m$ is consistently
estimated, (d) both $\g$ and $\m$ are inconsistently
estimated. Consistent estimators were obtained through the MLE in a
correctly specified parametric model, whereas inconsistent estimators
were obtained through the MLE in misspecified logistic regression
models that only included $X_1$.

\begin{figure}[H]
  \centerline{\includegraphics[width=4in]{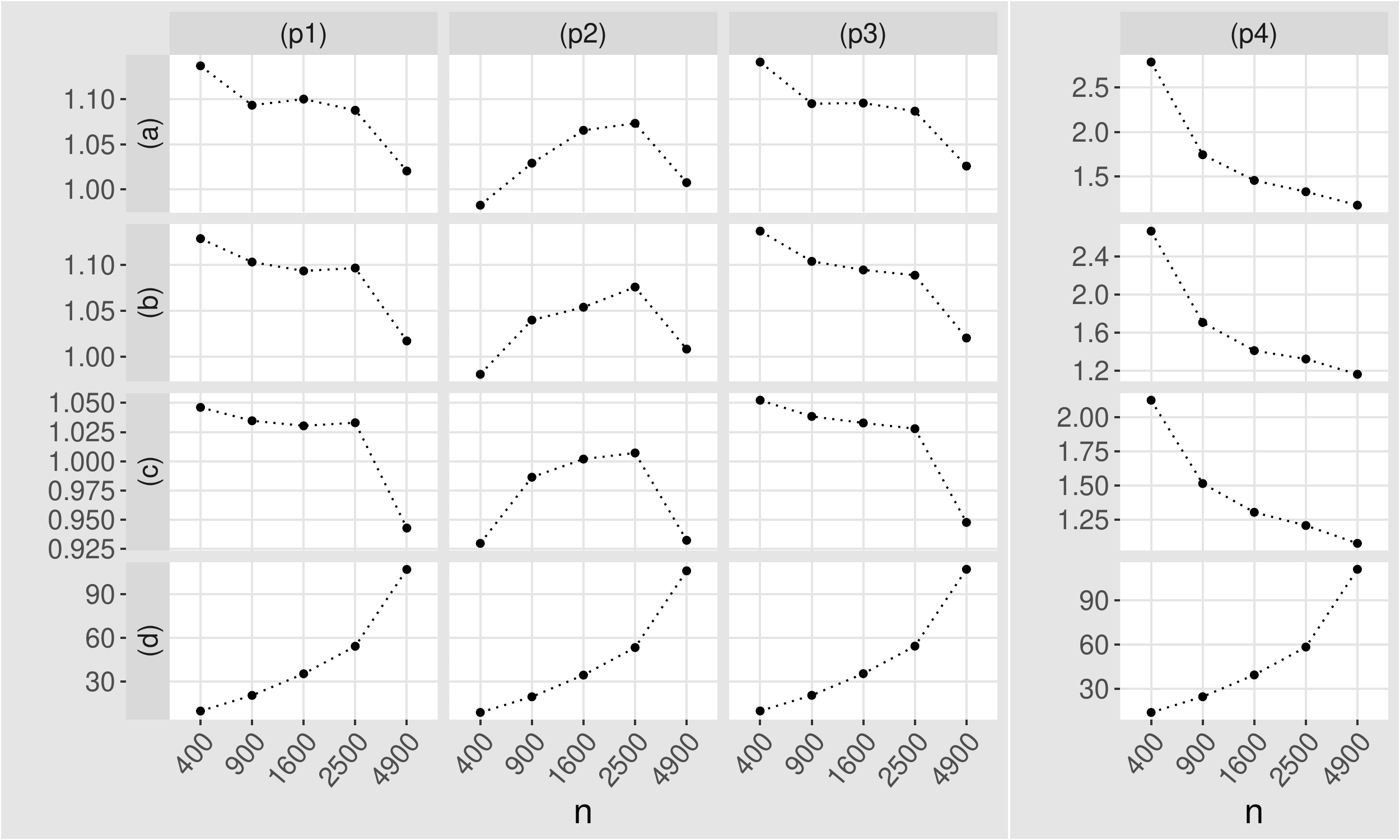}}
  \caption{Mean squared error of the posterior mean divided by the efficiency
    bound and scaled by $n$.}\label{fig:MSE}
\end{figure}

For the prior distribution we used a $\text{Beta}(\alpha, \beta)$
distribution. For each of scenarios in the above paragraph we consider
four different prior specifications: (p1) correct mean and large
variance, (p2) correct mean and small variance, (p3) incorrect mean
and large variance and (p4) incorrect mean and small variance. The
correct mean was computed as the true value of the parameter
$\theta= 0.77$, the incorrect mean was specified as
$\theta= 0.23$.  Large and small variances are
$\sigma^2=  0.018$. and
$\sigma^2= 0.16$, respectively. The Beta distribution
was reparameterized through
\[\alpha =  \theta\left(\frac{\theta(1 - \theta)}{\sigma^2} - 1\right);\quad
  \beta = \alpha\left(\frac{1}{\theta} - 1\right)\] We evaluate the
performance of the posterior mean as an estimator of $\theta$ in terms
of: (i) relative efficiency defined as the ratio of mean squared error
and and efficiency bound scaled by $n$, (ii) coverage of the 95$\%$
confidence intervals and, (iii) absolute bias scaled by $n^{1/2}$.
The results for the MSE and coverage are presented in Figure
\ref{fig:MSE} and Figure \ref{fig:Cov95}, respectively.  The results
for bias are presented in Figure \ref{fig:bias} in the Supplementary
Materials.

The above plot shows that the MSE is comparable for prior
specifications (p1)-(p3), which reflect scenarios with uncertainty
about the parameter value, or scenarios with certainty about a correct
value. In contrast, the MSE for prior specification (p4), which
illustrates a scenario with prior certainty about an incorrect value,
is generally larger and is shown in a different scale. Likewise,
scenarios (a)-(c), in which at least one of the nuisance parameters
$\m$ and $\g$ is consistently estimated, yield mean squared errors
which converge at $n$-rate. According to Theorems~\ref{theo:aslin} and
\ref{theo:ber}, it is expected that the $n$-scaled mean squared error
for scenario (a) will converge to the efficiency bound. This is
illustrated in the simulation. The fact that the same convergence
seems to hold for (b) and (c) may be an artifact of this particular
data generating mechanism. In contrast, as also expected, the MSE for
scenario (d) diverges at $n$-rate, reflecting on the inconsistency of
the estimator. This is a consequence of a diverging bias (Figure
\ref{fig:bias} in the Supplementary Materials).
\begin{figure}[H]
  \centerline{\includegraphics[width=4in]{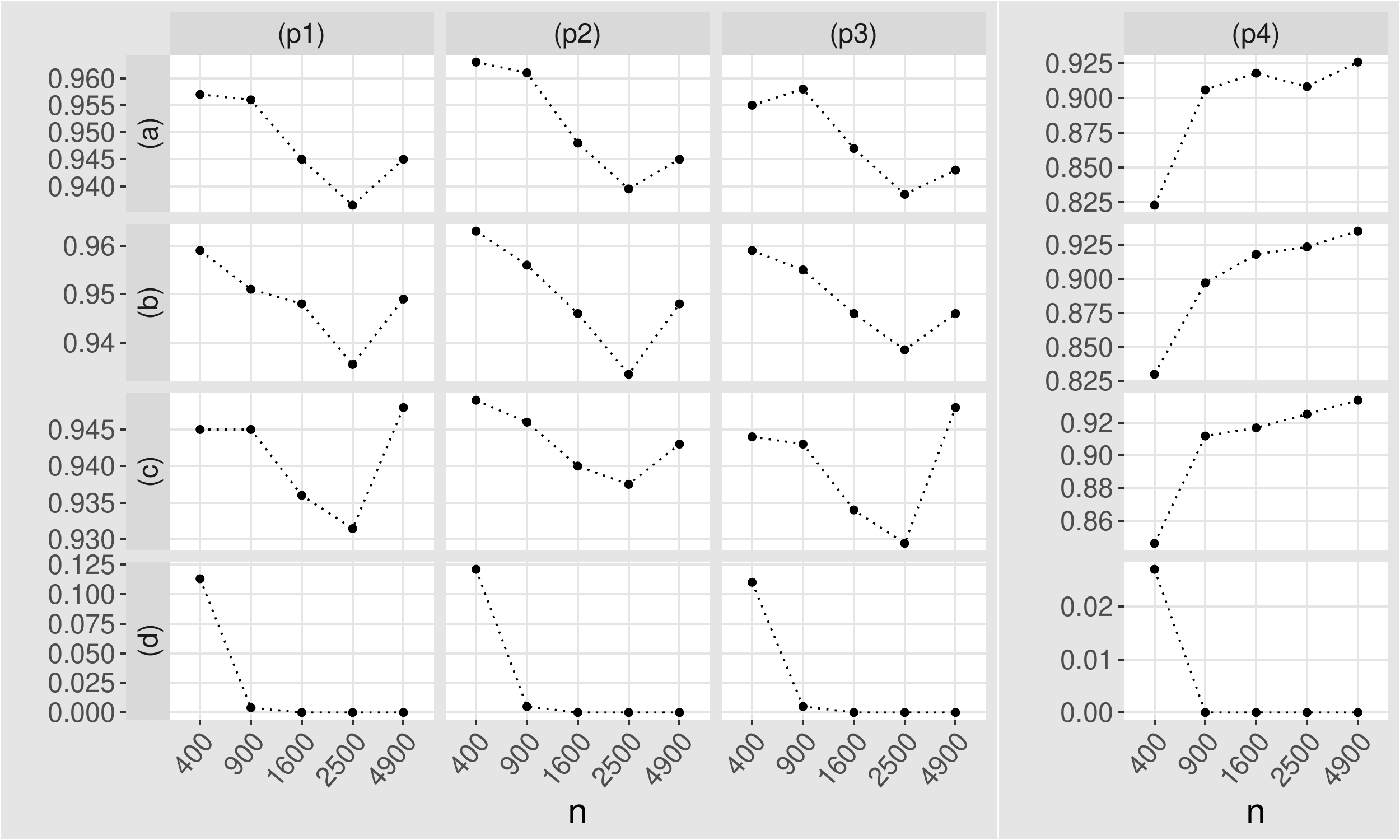}}
  \caption{Coverage of 95\% credible intervals.}\label{fig:Cov95}
\end{figure}
The coverage of credible intervals also seems to converge to the
nominal value in all scenarios, except under inconsistent estimation
of both nuisance parameters (d). The convergence seems to be much
slower for the case of an incorrectly centered but highly precise
prior (p4).


\section{Discussion and extensions}

Theorem~\ref{theo:aslin} is the basic asymptotic linearity result for
targeted MLE-type estimators. Recent developments in this literature
have showed that it is possible to construct targeted MLEs with
several additional properties. For instance, recent manuscripts
\citep{van2014targeted, benkeser2016doubly, diaz2017doubly,
  diaz2019statistical} have studied methods to construct targeted MLEs
that relax assumption \label{ass:rate} in the sense that the resulting
estimator converges to a Gaussian variable even when one at most of
the nuisance parameters is inconsistently estimated. Similarly,
\cite{gruber2012targeted,Colantuoni2015, Diaz2016, diaz2018improved}
have proposed targeted ML estimators with the additional property that
they outperform a given estimator in terms of asymptotic variance.
Methods to endow the TMLE with such additional properties generally
operate by adding additional auxiliary covariates to the submodels in
(\ref{eq:submodel}).  The covariates are carefully constructed to
ensure that the resulting least favorable submodel contains the
appropriate score functions, and thus the targeted MLE $\tilde\eta$ is
enhanced to solve certain estimating score equations that endow the
targeted MLE with the above properties. The Bayesian procedure we
propose can also be endowed with such properties, via the Bernstein
von-Mises result in Theorem~\ref{theo:ber}. This may be done by using
the corresponding enhanced TMLE $\tilde\eta$ in the construction of
the likelihood function for the Bayes procedure (e.g., equation
(\ref{eq:likeli})). Theorem~\ref{theo:ber} implies that the posterior
distribution will also inherit the additional properties of the
targeted MLE. This may have important applications for constructing
adjusted Bayesian estimators in adaptive sequential designs, and
Bayesian estimators with the property that their asymptotic variance
is never smaller than the variance of an unadjusted estimator.


\bibliographystyle{plainnat}
\bibliography{refs}

\end{document}